\documentclass[aps,prb,preprint,eqsecnum,titlepage]{revtex4-2}

\usepackage{amsmath,amssymb}
\usepackage{graphicx}
\usepackage{xcolor}
\counterwithout{equation}{section}
\usepackage[
  colorlinks=false,
  citebordercolor=green,
  linkbordercolor=green,
  urlbordercolor=green,
  pdfborder={0 0 1}
]{hyperref}

\begin{document}
\title{The First Variational Formula and the Ostrogradsky Formalism}

\author{Drew Watson, Matthew Pontius, Charles Torre}
\affiliation{Department of Physics, Utah State University, Logan, Utah 84322-4415, USA}
\date{June 2026}         

\begin{abstract}

We present a derivation at a level suitable for undergraduates of the Ostrogradsky formalism for Lagrangians in classical mechanics that depend upon an arbitrary number of time derivatives of the configuration. From the boundary term in the first variation of the Lagrangian we derive the Ostrogradsky formulas that define the Hamiltonian formulation of mechanical systems. Worked examples, exercises, and applications to the literature are also provided. An accompanying computer program that implements the formalism is discussed in the Supplementary Materials, and code for computing Hamiltonians via the Ostrogradsky formalism is provided in the Supplementary Materials and in a GitHub repository.

\end{abstract}

\maketitle

\section{Introduction}
\label{intro}

This article is concerned with the Lagrangian and Hamiltonian formulations of classical mechanics in cases where the Lagrangian depends upon accelerations or higher derivatives of the configuration variables. The variation of the Lagrangian defines both the Euler-Lagrange expression and  a boundary term.  As discussed in classical mechanics textbooks, the Euler-Lagrange equation is used to define differential equations of motion, while the boundary term determines the boundary conditions for the associated variational principle and is featured in the connection between symmetries and conservation laws \cite{Goldstein, LandauLifshitz, Lanczos, Whittaker}. It has been known for some time in the physics research literature (usually pertaining to field theories) that the boundary term can also be used to define the Hamiltonian formulation of dynamics \cite{Torre, Barnich, Joohan, CrnkovicWitten}. In fact, much of the interest surrounding Lagrangians depending on higher order derivatives originates in the gravity literature \cite{Woodard_2007, HigherOrderGravity, Pavsic}. 

Our aim in this work is to present this feature of the Lagrangian formulation of mechanics at a level suitable for presenting to undergraduates and, in particular, to use it to make more transparent the construction of the Hamiltonian formulation for Lagrangians that depend upon accelerations or higher derivatives of the configuration variables. We also aim to make explicit a point often left implicit in undergraduate classical mechanics: the canonical momenta can be read from the boundary term in the first variation of the Lagrangian. This observation is almost hidden in the ordinary first-order case, where it simply reproduces $p=\partial L/\partial \dot q$, but it becomes especially useful for deriving the less obvious Ostrogradsky momenta in higher-derivative mechanics. This construction for higher derivatives is due to Ostrogradsky, at least in the case where the highest derivatives appear in the Lagrangian in a suitably non-degenerate manner \cite{Ostrogradsky, Woodard}. We do not deal in this article with degenerate Lagrangians, for which the so-called Dirac-Bergmann algorithm must be employed. The reader is encouraged to consult \cite{Dirac, Brown_2022, Brown_2023} if they wish to learn more about this approach. 

Ostrogradsky first discussed this in \cite{Ostrogradsky}. We seek to present a derivation that is more modern, streamlined, and accessible when compared to Ostrogradsky's original work. While Woodard explains the Ostrogradsky formalism in \cite{Woodard}, to our knowledge no derivation of Ostrogradsky's construction exists in the modern literature which emphasizes the role of the boundary term in the variation of $L$. Here we will show that Ostrogradsky's construction, which might appear rather clever and somewhat {\it ad hoc}, can be understood quite simply in terms of the boundary term in the first variational formula. We share numerous examples, insights, and extensions to the literature to demonstrate the formalism's broad applicability. In addition, we supply computer code for calculating Hamiltonians for Lagrangians depending on higher order derivatives in the Supplementary Material and on GitHub\footnote{\url{https://github.com/Drew-Watson-117/hamiltonian}}.

In the next section we summarize the Ostrogradsky construction. In Section \ref{section1} we explain why the boundary term arising from the variation of the Lagrangian can be used to define canonical coordinates and momenta. In Section \ref{section2} we show that the Ostrogradsky formulas emerge from calculating the boundary term for an arbitrary Lagrangian (albeit with only one dependent variable). In the Supplementary Materials, we discuss a computer algorithm for the Ostrogradsky formalism and provide implementations of the algorithm in Python and Maple.

\noindent\section{Summary of Ostrogradsky's Construction}

\subsection{Ordinary Hamiltonian Mechanics}

We begin by briefly reviewing the familiar form of the Hamiltonian formulation of mechanics. For simplicity, we restrict to a single degree of freedom. Consider a Lagrangian $L=L(q,\dot{q}, t)$, which is a function of a generalized coordinate $q$, its first time derivative $\dot q$, and the time $t$. The Euler-Lagrange equations of motion for $q=q(t)$ are given by
\begin{equation}
\frac{\partial L}{\partial q} - \frac{d}{dt}\frac{\partial L}{\partial \dot q}=0.
\label{EL0}
\end{equation}
The Hamiltonian formulation of this second-order ordinary differential equation is as follows. The canonical coordinates and momenta $(Q, P)$ are defined by 
\begin{equation}
    Q=q
    \label{Q0}
\end{equation}
    
\begin{equation}
    P=\frac{\partial L}{\partial\dot{q}}.
    \label{P0}
\end{equation}

\noindent where the latter equation defines $P$ as the canonical pairing to $Q$. The Hamiltonian $H=H(Q,P)$ is defined by

\begin{equation}
    H=P\dot{q}-L,
\end{equation}
where it is understood that one has eliminated $(q, \dot q)$ in favor of $(Q, P)$ using (\ref{Q0}), (\ref{P0}). For this to be possible we require $P$ to depend non-trivially on $\dot q$:
 \begin{equation}
 \frac{\partial}{\partial\dot q}\left(\frac{\partial L}{\partial \dot q}\right) \neq 0.
 \end{equation}

Hamilton's equations for curves in phase space, $Q=Q(t)$, $P=P(t)$,  are given by 

\begin{equation}
    \dot{Q}=\frac{\partial H}{\partial P}
\end{equation}

\begin{equation}
    \dot{P}=-\frac{\partial H}{\partial Q}.
\end{equation}
 These 2 first-order ordinary differential equations are equivalent to the Euler-Lagrange equation (\ref{EL0}).

\subsection{Ostrogradsky's Construction}
\label{subsection}

We now  allow the Lagrangian to depend upon higher time derivatives of $q$,
\begin{equation}
L=L(q,q^{(1)},q^{(2)},\dots,q^{(N)}, t),
\end{equation}
where  $q^{(k)}$ denotes the $k$th time derivative of $q$.  The Euler-Lagrange equation is \cite{Goldstein}
\begin{equation}
\frac{\partial L}{\partial q} - \frac{d}{dt}\frac{\partial L}{\partial q^{(1)}} + \frac{d^2}{dt^2}\frac{\partial L}{\partial q^{(2)}} + \dots + 
(-1)^N\frac{d^N}{dt^N}\frac{\partial L}{\partial q^{(N)}} = 0.
\label{ELgen}
\end{equation}
The Euler-Lagrange equation is of order $2N$ (at most), which implies the requirement of $2N$ initial conditions to specify the motion and the existence of $2N$ canonical coordinates. Ostrogradsky's Hamiltonian formulation for such Lagrangians is as follows.
Define the canonical coordinates
\begin{equation}\label{eqn:canonical_coordinates}
Q^1= q, \quad Q^2 = \dot{q},\quad  Q^3 = \ddot{q},\quad  \dots, \quad  Q^{N} = q^{(N-1)}.
\end{equation}
The conjugate momenta are defined by
\begin{equation}\label{eqn:canonical_momenta}
    P_i\equiv\sum\limits_{j=i}^{N}(-\frac{d}{dt})^{j-i}\frac{\partial L}{\partial q^{(j)}}, \quad i = 1,\dots, N.
\end{equation}
Note that for $N=1$ the definitions above reduce to the familiar canonical pairs for Hamiltonian mechanics. We assume that the Lagrangian depends non-degenerately on $q^{(N)}$, 
\begin{equation}
\frac{\partial}{\partial q^{(N)}}\left(\frac{\partial L}{\partial q^{(N)}} \right) \neq 0,
\label{Onondegen}
\end{equation}
meaning we can solve for $q^{(N)}$ in terms of $Q^1,\dots,Q^{N}$ and $P_N$ :
\begin{equation}
q^{(N)}=A(Q^1,Q^2,\dots,Q^{N},P_N).
\end{equation} The Hamiltonian is then defined by 
\begin{equation}
    H=\sum\limits_{i=1}^{N-1} P_i Q^{i+1} + P_N A -L(Q^1,\dots, Q^N, A). \label{hamiltonian_one_dep_var}
\end{equation}
Taking the differential of Equation \ref{hamiltonian_one_dep_var} and setting it equal to \(dH(Q^i,P_i,A(Q^i,P_i),t)\), the Hamilton equations are obtained:
\begin{equation}
    \dot{Q^i}=\frac{\partial H}{\partial P_i}, \label{hamiltons_eqn_pos}
\end{equation}
and
\begin{equation}
    \dot{P_i}=-\frac{\partial H}{\partial Q^i}. \label{hamiltons_eqn_mom}
\end{equation}
These $2N$ first-order equations are equivalent to the single Euler-Lagrange equation (\ref{ELgen})\cite{Woodard}, which -- given (\ref{Onondegen}) -- is a differential equation of order $2N$. In fact, we can recover the Euler-Lagrange equation by using equations \ref{eqn:canonical_coordinates} and \ref{eqn:canonical_momenta} to substitute the canonical coordinates $Q^i,P_i$ for the configuration coordinate $q$ and its derivatives in the Hamilton equations and substituting one into the other.

\subsection{An example in the case \texorpdfstring{$N=2$}{N=2}: A perturbed harmonic oscillator}

Consider the second-order Lagrangian $L=L(q,\dot{q},\ddot{q})$ given by 

\begin{equation}
    L = -\frac{\epsilon m}{2 \omega^2} \ddot{q}^2 + \frac{m}{2} \dot{q}^2 - \frac{m \omega^2}{2} q^2. \label{example_lagrangian}
\end{equation}

Equation \ref{example_lagrangian} represents a Lagrangian for a simple harmonic oscillator which has been perturbed by a small dimensionless parameter $\epsilon$. Note that this Lagrangian satisfies the non-degeneracy condition i.e. that $\frac{\partial^2 L}{\partial \ddot{q}^2} \neq 0$. Using the definitions of the canonical coordinates and momenta given in Section \ref{subsection}, we obtain the following:

\begin{equation}
    Q^1=q \label{pos_1}
\end{equation}

\begin{equation}
    Q^2=\dot{q}
\end{equation}

\begin{equation}
    P_1=\frac{\partial L}{\partial \dot{q}}-\frac{d}{dt} \frac{\partial L}{\partial \ddot{q}}=m\dot{q}+\frac{\epsilon m}{\omega^2} \dddot{q}
\end{equation}

\begin{equation}
    P_2=\frac{\partial L}{\partial \ddot{q}}=-\frac{\epsilon m}{\omega^2} \ddot{q}. \label{last_momentum}
\end{equation}

Solving \ref{last_momentum} for $\ddot{q}$, we find that $A\equiv\ddot{q}=-\frac{\omega^2}{\epsilon m}P_2$. Substituting eqs. \ref{pos_1}--\ref{last_momentum} into \ref{hamiltonian_one_dep_var}, we arrive at

\begin{align}
     H(Q^1,Q^2,P_1,P_2)&=P_1Q^2+P_2A(Q^1,Q^2,P_2)-L(Q^1,Q^2,A(Q^1,Q^2,P_2)) \nonumber \\
     &=P_1Q^2-\frac{\omega^2}{2\epsilon m}P_2^2-\frac{m}{2}(Q^2)^2+\frac{m \omega^2}{2}(Q^1)^2. \label{(16)}
\end{align}

Using \ref{hamiltons_eqn_pos} and \ref{hamiltons_eqn_mom}, we may write Hamilton's equations

\begin{equation}
    \dot{Q^1}=\frac{\partial H}{\partial P_1}=Q^2
\end{equation}

\begin{equation}
    \dot{Q^2}=\frac{\partial H}{\partial P_2}=-\frac{\omega^2}{\epsilon m}P_2
\end{equation}

\begin{equation}
    \dot{P_1}=-\frac{\partial H}{\partial Q^1}=-m\omega^2Q^1
\end{equation}

\begin{equation}
    \dot{P_2}=-\frac{\partial H}{\partial Q^2}=-P_1+mQ^2.
\end{equation}

\subsection{An Example from the Gravity Literature: The Pais--Uhlenbeck Oscillator}

In attempts to construct renormalizable quantum field theories of gravity, one is naturally led to consider actions containing higher-derivative terms. A simple mechanical model that captures some of the resulting features is the Pais--Uhlenbeck oscillator, with Lagrangian \cite{Pavsic}
\begin{equation}
L(x,\dot x,\ddot x)=\frac12\left[\ddot{x}^2-(\omega_1^2+\omega_2^2)\dot{x}^2+\omega_1^2\omega_2^2 x^2\right],
\label{PU_L}
\end{equation}
optionally supplemented by a self-interaction term $-\frac{\Lambda}{4}x^4$. Higher-order terms also appear in attempts to improve the ultraviolet behavior of quantum field theories, including higher-curvature modifications of gravity. The price of introducing such terms is that the Hamiltonian formulation may acquire additional degrees of freedom, including modes with negative energy. The Pais–Uhlenbeck oscillator is useful because it is a simple mechanical model in which these issues can be studied explicitly.

The Euler-Lagrange equation following from \eqref{PU_L} is
\begin{equation}
x^{(4)}
+
(\omega_1^2+\omega_2^2)\ddot{x}
+
\omega_1^2\omega_2^2 x
=
0,
\end{equation}

or in factored form:

\begin{equation}
\left(
\frac{d^2}{dt^2}+\omega_1^2
\right)
\left(
\frac{d^2}{dt^2}+\omega_2^2
\right)x
=
0.
\end{equation}
Thus, when $\omega_1$ and $\omega_2$ are distinct positive frequencies, the free Pais--Uhlenbeck oscillator has solutions that are linear combinations of ordinary oscillatory terms,
\begin{align}
x(t)
=
A_1\cos(\omega_1 t)
+
B_1\sin(\omega_1 t)
+
A_2\cos(\omega_2 t)
+
B_2\sin(\omega_2 t).
\label{PU_general_solution}
\end{align}
This makes the model physically instructive: at the level of the free equation of motion it resembles two ordinary harmonic oscillators, but the Hamiltonian formulation reveals that one of these modes carries the opposite sign of energy. The Pais--Uhlenbeck oscillator therefore provides a simple example in which bounded oscillatory motion and the Ostrogradsky instability can be discussed in the same system.

This is a second-order ($N=2$) Lagrangian, so the Ostrogradsky canonical coordinates are
\begin{equation}
Q^1=x,\qquad Q^2=\dot x.
\end{equation}
The conjugate momenta are
\begin{align}
P_2&=\frac{\partial L}{\partial \ddot x}=\ddot x,
\label{PU_P2}
\\
P_1&=\frac{\partial L}{\partial \dot x}-\frac{d}{dt}\frac{\partial L}{\partial \ddot x}
=-(\omega_1^2+\omega_2^2)\dot x-\dddot x.
\label{PU_P1}
\end{align}
The non-degeneracy condition holds since $\frac{\partial^2 L}{\partial \ddot x^2}=1\neq 0$, and we can solve
\begin{equation}
A\equiv \ddot x = P_2.
\end{equation}
Substituting into the definition
\begin{equation}
H=P_1Q^2+P_2A-L(Q^1,Q^2,A),
\end{equation}
we obtain
\begin{equation}
H(Q^1,Q^2,P_1,P_2)=P_1Q^2+\frac12P_2^2+\frac12(\omega_1^2+\omega_2^2)(Q^2)^2-\frac12\omega_1^2\omega_2^2(Q^1)^2.
\label{PU_H}
\end{equation}
If the interaction $\frac{\Lambda}{4}x^4$ is included in \eqref{PU_L}, then \eqref{PU_H} is modified by the additional term $+\frac{\Lambda}{4}(Q^1)^4$.

Hamilton's equations are therefore
\begin{align}
\dot Q^1&=\frac{\partial H}{\partial P_1}=Q^2,
\end{align}
\begin{align}
\dot Q^2&=\frac{\partial H}{\partial P_2}=P_2,
\end{align}
\begin{align}
\dot P_1&=-\frac{\partial H}{\partial Q^1}=\omega_1^2\omega_2^2 Q^1 \;\; \bigl(\text{or }\omega_1^2\omega_2^2 Q^1-\Lambda (Q^1)^3\bigr),
\end{align}
\begin{align}
\dot P_2&=-\frac{\partial H}{\partial Q^2}=-P_1-(\omega_1^2+\omega_2^2)Q^2. 
\end{align}

It is important to note that the momentum \(P_1\) appears linearly in the Hamiltonian. This is not accidental, but rather a general feature of the Ostrogradsky formalism, known as the Ostrogradsky instability theorem. For any non-degenerate higher-derivative Lagrangian, the resulting Hamiltonian is linear in at least one canonical momentum, and is therefore unbounded from below (and above).

Since \(P_1 \in (-\infty, \infty)\), the Hamiltonian can take arbitrarily large positive and negative values. This contrasts with the Hamiltonian of a free particle, \(H = P_1^2/2m\), which is unbounded above but bounded below, and thus admits a stable minimum energy state. In the Pais--Uhlenbeck oscillator, no such lower bound exists, and consequently the system has no ground state.

Upon diagonalization, the Pais--Uhlenbeck oscillator can be decomposed into two normal modes, one carrying positive energy and the other negative energy. As a result, the total Hamiltonian is not positive definite. While the free system admits bounded oscillatory solutions, the presence of both positive- and negative-energy modes allows for arbitrarily large energy transfer between them. In the presence of interactions or perturbations, this leads to runaway behavior, as the amplitudes of the modes can grow without bound while preserving total energy. A more detailed discussion of the origin and implications of the Ostrogradsky instability in the PU oscillator may be found in \cite{Smilga, Kovner, ajp_puo}.

\subsection{Ostrogradsky's Construction for Multiple Dependent Variables}

Here we briefly indicate what to do if there are multiple degrees of freedom. We suppose there are multiple degrees of freedom denoted by $q^\alpha$, with $\alpha = 1,2,\dots M$. When a Lagrangian depends on multiple dependent variables, one simply adds a summation term for each dependent variable. Label dependent variables $q^1,q^2,q^3,\dots,q^M$.  We define the canonical coordinates
\begin{equation}
    Q^{\alpha,i}= \frac{d^{i-1}}{d t^{i-1}} q^\alpha
\end{equation}
and momenta
\begin{equation}
    P_{\alpha,i}\equiv\sum\limits_{j=i}^{N_\alpha}(-\frac{d}{dt})^{j-i}\frac{\partial L}{\partial (q^{\alpha})^{(j)}}, \quad i = 1,\dots, N_\alpha.
    \label{2.7}
\end{equation}
where $\alpha$ sums over the dependent variables (e.g. $x,y,z$), $(q^{\alpha})^{(j)}$ is the $j$th time derivative of $q^\alpha$, and $N_\alpha$ denotes the highest-order derivative of $q^\alpha$ present in the Lagrangian. The Hamiltonian is then given by
\begin{equation}
    H=\sum\limits_{\alpha=1}^M\left(\sum\limits_{i=1}^{N_\alpha-1} P_{\alpha,i}Q^{\alpha,i+1} + P_{\alpha,N_\alpha}A^\alpha\right) - L , \label{eqn:hamiltonian_multiple_dependent}
\end{equation}
where as in \ref{subsection} we have
\begin{equation}
    (q^\alpha)^{(N_{\alpha})} = A^\alpha(Q^{1,1}, Q^{1,2}, \dots, Q^{M,N_M}, P_{\alpha,N_{\alpha}})
\end{equation}
Hamilton's equations are then defined by

\begin{equation}
    \dot{Q^{\alpha,i}}=\frac{\partial H}{\partial P_{\alpha,i}}
\end{equation}
and
\begin{equation}
    \dot{P_{\alpha,i}}=-\frac{\partial H}{\partial Q^{\alpha,i}}.
\end{equation}

\section{Defining the canonical pairs via the boundary term}
\label{section1}

In ordinary undergraduate mechanics, the canonical momentum is usually introduced by the formula
\begin{equation}
p=\frac{\partial L}{\partial \dot q},
\end{equation}
and its usefulness is then justified by the Legendre transform and Hamilton's equations. For first-order Lagrangians this definition is simple enough that one may easily take it as given. However, from the variational point of view, the formula has a more structural origin: it is precisely the coefficient of \(\delta q\) in the boundary term of the first variation,
\begin{equation}
\delta L
=
\left(
\frac{\partial L}{\partial q}
-
\frac{d}{dt}\frac{\partial L}{\partial \dot q}
\right)\delta q
+
\frac{d}{dt}\left(
\frac{\partial L}{\partial \dot q}\,\delta q
\right).
\end{equation}
Thus the boundary term contains the canonical pairing \(p\,\delta q\). In fact, the canonical coordinates and momenta can be read directly from this boundary term. In the ordinary case this reproduces the familiar definition of momentum, while in the higher-derivative case it provides a systematic derivation of the Ostrogradsky momenta, as shown in Section \ref{section2}.

We now make this observation more precise from the perspective of the Hamiltonian formalism and establish notation. A Hamiltonian system may be defined by a collection of canonical pairs---coordinates and momenta---denoted by $(Q^i,P_i)$, $i=1,\dots,N$, together with a function, the Hamiltonian $H=H(Q,P)$. For simplicity, we restrict attention to Lagrangians and Hamiltonians with no explicit time dependence.
The Hamilton equations are the Euler-Lagrange equations associated to the phase space Lagrangian,
\begin{equation}
\mathbf{L}(Q,P,\dot Q)=P_i\dot Q^i-H(Q,P),\label{eqn:phase_space_lagrangian}
\end{equation}
where we are using the Einstein summation convention and dots denote time derivatives. The Hamilton equations can be obtained by writing the variation of this Lagrangian in the following form:
\begin{equation}
\delta\mathbf{L} = \left(\dot Q^i - \frac{\partial H}{\partial P_i}\right)\delta P_i + \left(-\dot P_i - \frac{\partial H}{\partial Q^i}\right)\delta Q^i + \frac{d}{dt} \left(P_i \delta Q^i\right)
\end{equation}
We call the last term, $\frac{d}{dt} \left(P_i \delta Q^i\right)$, the ``boundary term'' since it defines endpoint terms in the variation of the action integral, $S = \int L dt$.

The configuration space Lagrangian can be written as 
\begin{equation}
    \mathcal{L} =\mathcal{L}(q^i,\dot q^i),
\end{equation}
with variation
\begin{equation}
    \delta{\cal L} = \left(\frac{\partial\cal L}{\partial q^i}-\frac{d}{dt}\frac{\partial\cal L}{\partial\dot q^i}\right)\delta q^i+\frac{d}{dt}\left(\frac{\partial\cal L}{\partial\dot q^i}\delta q^i\right).
\end{equation}
To find the canonical pairs associated with $\cal L$, we demand that
\begin{equation}\label{eqn:lagrangians_equal}
{\cal L}(q^i, \dot q^i) = \mathbf{L}\left(Q^i(q^i,\dot q^i), P_i(q^i, \dot q^i)\right).
\end{equation}
The equation above implies that the variations of the two Lagrangians must also be equal,
\begin{equation}
\delta\cal L = \delta\mathbf{L}.
\end{equation}
Substituting the explicit variations into the equation above gives
\begin{equation}
    \left(\frac{\partial\cal L}{\partial q^i}-\frac{d}{dt}\frac{\partial\cal L}{\partial\dot q^i}\right)\delta q^i+\frac{d}{dt}\left(\frac{\partial\cal L}{\partial\dot q^i}\delta q^i\right) = \left(\dot Q^i - \frac{\partial H}{\partial P_i}\right)\delta P_i + \left(-\dot P_i - \frac{\partial H}{\partial Q^i}\right)\delta Q^i + \frac{d}{dt} \left(P_i \delta Q^i\right),
\end{equation}
and invoking the Euler-Lagrange equations and Hamilton's equations results in
\begin{equation}
    \frac{d}{dt}\left(\frac{\partial\cal L}{\partial\dot q^i}\delta q^i\right) = \frac{d}{dt} \left(P_i \delta Q^i\right).
\end{equation}
The equation above is only satisfied if the canonical pair is defined as
\begin{equation}
    (Q^i,P_i)=\left(q^i,\frac{\partial\cal L}{\partial \dot{q}^i}\right).
\end{equation}
This demonstrates that the boundary term of the Lagrangian (in configuration space) defines the canonical pair in phase space.

\section{Deriving Ostrogradsky's Canonical Pairs from the Boundary Term}
\label{section2}
We now consider the variation of a general Lagrangian, viewed as a function of configuration space variables and their time derivatives to any order,  thereby establishing the form of the Euler-Lagrange expression and the boundary term.  Consider a Lagrangian which depends upon derivatives of the dependent variable $q$ up to order $N$
\begin{equation}
L = L(q, q^{(1)}, q^{(2)}, \dots,q^{(N)}, t).
\end{equation}
  For ease of reading, we will only use a single dependent variable -- the superscript indicates the number of time derivatives. We  leave it as an exercise for the reader to repeat this derivation for an arbitrary number of dependent variables.

Let $F = F(q^{(1)},\dots, q^{(N)}, t)$. It is straightforward to show by integration by parts that, for any $k>0$, 
\begin{equation}
\label{lemma1}
F \delta q^{(k)} = (-1)^k \frac{d^k F}{dt^k}\delta q + \frac{d}{dt}\left(\sum_{j=0}^{k-1} (-1)^{j}\frac{d^{j}F}{dt^{j}}\delta q^{(k-j-1)}\right).
\end{equation}

For any $N\times N$ array $a_{ij}$ we have
\begin{equation}\label{lemma2}
\sum_{j=1}^N\sum_{i=1}^j a_{ji}=\sum_{i=1}^N\sum_{j=i}^N a_{ji} ,
\end{equation}
since explicitly summing shows that each side is the sum over all the $a_{ji}$ where $i\leq j$.

We now show that the variation $\delta L$ of the Lagrangian takes the form,
\begin{equation}\label{eqn:begin_thm_1}
\delta L = {\cal E}(L)\delta q + \frac{d}{dt} \Theta(\delta q),
\end{equation}
where
\begin{equation}
{\cal E}(L) = \sum_{k = 0}^N (-1)^k\frac{d^k}{dt^k} \frac{\partial L}{\partial q^{(k)}},
\end{equation}
and the boundary terms
\begin{equation}\label{eqn:end_thm_1}
\Theta(\delta q) = \sum_{i=1}^N\sum_{j=i}^N \left[\left(-\frac{d}{dt}\right)^{j-i} \frac{\partial L}{\partial q^{(j)}}\right] \delta q^{(i-1)}.
\end{equation}

The variation of $L$ is 
\begin{equation}
\delta L = \sum_{k=0}^N \frac{\partial L}{\partial q^{(k)}}\delta q^{(k)}.
\end{equation}
Now apply Equation \ref{lemma1} with $F = \frac{\partial L}{\partial q^{(k)}}$:
\begin{equation}
\delta L = \sum_{k=0}^N (-1)^k \frac{d^k }{dt^k}\frac{\partial L}{\partial q^{(k)}}\delta q +  \frac{d}{dt}\sum_{k=1}^N  \left(\sum_{j=0}^{k-1} (-1)^{j}\frac{d^{j}}{dt^{j}}\frac{\partial L}{\partial q^{(k)}}\delta q^{(k-j-1)}\right)
\end{equation}
Define $i = k - j $.  The summation in the second term can be written
\begin{equation}
\sum_{k=1}^N  \left(\sum_{j=0}^{k-1} (-1)^{j}\frac{d^{j}}{dt^{j}}\frac{\partial L}{\partial q^{(k)}}\delta q^{(k-j-1)}\right)
= \sum_{k=1}^N\left(\sum_{i=1}^k (-1)^{k-i}\frac{d^{k-i}}{dt^{k-i}}\frac{\partial L}{\partial q^{(k)}}\delta q^{(i-1)}\right).
\end{equation}
Relabel $k=j$ in the second term to get
\begin{equation}
\delta L = \sum_{k=0}^N (-1)^k \frac{d^k }{dt^k}\frac{\partial L}{\partial q^{(k)}}\delta q + \frac{d}{dt} \left(\sum_{j=1}^N\sum_{i=1}^j \left(-\frac{d}{dt}\right)^{j-i}\frac{\partial L}{\partial q^{(j)}}\delta q^{(i-1)}\right).
\end{equation}
Finally, apply Equation \ref{lemma2} to the second term to get
\begin{equation}
\delta L = \sum_{k=0}^N (-1)^k \frac{d^k }{dt^k}\frac{\partial L}{\partial q^{(k)}}\delta q + \frac{d}{dt}\left(\sum_{i=1}^N \sum_{j=i}^N \left(-\frac{d}{dt}\right)^{j-i}\frac{\partial L}{\partial q^{(j)}}\delta q^{(i-1)}\right).
\end{equation}

As discussed in Section \ref{section1}, the boundary term $\Theta$ is of the form $P_i\, \delta Q^i$,  where (in the case of a non-degenerate Lagrangian) $Q^i$ and $P_i$ are canonical pairs (it is left to the reader to extrapolate the argument made in Section \ref{section1} to the more general case). We can therefore identify the definitions of the canonical coordinates and momenta in terms of the variables $(q, \dot q, \ddot q, q^{(3)}, q^{(4)}, \dots)$:
\begin{equation}
Q^i = q^{(i-1)},\quad P_i = \sum_{j=i}^N \left[\left(-\frac{d}{dt}\right)^{j-i} \frac{\partial L}{\partial q^{(j)}}\right]. 
\end{equation}
These definitions are precisely those of Ostrogradsky, as reported by Woodard \cite{Woodard}

Furthermore, it is possible to rewrite the first variational formula purely in terms of Euler-Lagrange operations.  For any function $F = F(t, q, q^{(1)}, q^{(2)}, \dots, q^{(N)})$ and a chosen variable $q^{(s)}$, we define the corresponding Euler-Lagrange operation (``variational derivative''):
\begin{equation}
\label{ELdef}
\frac{\delta F} {\delta q^{(s)}} = \frac{\partial F}{\partial q^{(s)}} - \frac{d}{dt} \frac{\partial F}{\partial q^{(s+1)}} + \frac{d^2}{dt^2} \frac{\partial F}{\partial q^{(s+2)}} - \dots = \sum_{i=0}^{N-s}\left(-\frac{d}{dt} \right)^i\frac{\partial F}{\partial q^{(s+i)}}.
\end{equation}
When $s=0$ this is the usual Euler-Lagrange expression. With this definition and using Equations \ref{eqn:begin_thm_1}-\ref{eqn:end_thm_1} we get that the first variational formula for $L$ is given by
\begin{equation}
\delta L = \frac{\delta L}{\delta q}{\delta q}+ \frac{d}{dt}\Theta, 
\end{equation}
where
\begin{equation}
\Theta = \sum_{i=0}^{N-1} \frac{\delta L}{\delta q^{(i+1)}}\delta q^{(i)}.
\end{equation}
This is consistent with the more general results from classical field theory  \cite{Barnich}.  We note that $\Theta$ will depend upon as many as $2N-1$ derivatives of $q$.  Using the Euler-Lagrange operation defined in (\ref{ELdef}), the canonical momenta are defined by the elegant formula
\begin{equation}
P_i = \frac{\delta L}{\delta q^{(i)}}.
\end{equation}

The boundary term also explains the relation to the \(2N\) initial data. Since the Euler--Lagrange equation is generically of order \(2N\), a solution requires \(2N\) initial conditions. The boundary term contains variations of
\[
q,\ \dot q,\ \dots,\ q^{(N-1)},
\]
which are therefore the natural \(N\) configuration variables. Their coefficients in the boundary term give the corresponding \(N\) conjugate momenta. Thus the boundary term organizes the \(2N\) initial data into \(N\) canonical pairs.

\subsection{Examples}
\label{section2a}

We consider the familiar case where the Lagrangian
only depends upon the generalized coordinates and their velocities ({\it i.e.}, $N=1$).  We then have
\begin{equation}
{\cal E}(L) = \frac{\delta L}{\delta q} = \frac{\partial L}{\partial q} - \frac{d}{dt}\frac{\delta L}{\delta q^{(1)}},\quad
\Theta = \frac{\delta L}{\delta q^{(1)}} \delta q = \frac{\partial L}{\partial q^{(1)}} \delta q.
\end{equation}
Here we see that the canonical coordinates and momenta are defined by 
\begin{equation}
Q^1 = q,\quad  P_1 = \frac{\partial L}{\partial q^{(1)}},
\end{equation}
 in agreement with the standard canonical formalism.
 
Next we consider $N=2$:
\begin{equation}
\Theta = \frac{\delta L}{\delta q^{(1)}} \delta q + \frac{\delta L}{\delta q^{(2)}}\delta q^{(1)} = \left(\frac{\partial L}{\partial q^{(1)}} - \frac{d}{dt} \frac{\partial L}{\partial q^{(2)}}\right)\delta q + \frac{\partial L}{ \partial q^{(2)}}\delta q^{(1)}, 
\end{equation}
so that canonical pairs are defined by
\begin{equation}
Q^1 = q,\quad P_1 = \frac{\partial L}{\partial  q^{(1)}} - \frac{d}{dt} \frac{\partial L}{\partial  q^{(2)}},
\end{equation}
and
\begin{equation}
Q^2 = q^{(1)},\quad P_2 = \frac{\partial L}{\partial  q^{(2)}}.
\end{equation}
This agrees with the Ostrogradsky formalism, as reported, {\it e.g.}, in Woodard's article \cite{Woodard}.

\section{Discussion}

A detailed derivation of the form of Ostrogradsky's canonical coordinates is missing from the modern literature. While well-known references such as Woodard explain Ostrogradsky's construction, no reasoning is given for the form of Ostrogradsky's canonical coordinates \cite{Woodard}. In this work, we have shown that the Ostrogradsky formalism arises from calculating the boundary term in the variation of the Lagrangian. This work has presented the Ostrogradsky formalism and the idea of defining the conjugate pairs emphasizing the boundary term in a novel form that is suitable for undergraduate physics students. In the Supplementary Materials we discuss a novel code which implements this formalism. Deeper insight into the Ostrogradsky formalism will benefit researchers in multiple areas of physics, particularly quantum gravity, but also addresses a question often left unaddressed in many undergraduate classical mechanics courses, namely how to deal with Lagrangians containing higher derivatives. A better understanding of this subject can help enhance research in areas which feature Lagrangians depending on higher-order derivatives, and serves as an excellent launching point to further studies in the theory, such as the Dirac-Bergmann algorithm and Ostrogradsky instability mentioned previously.

\section{Statement of Author Roles}

All authors made significant contributions to the work outlined in this manuscript. During the research, Mr. Watson focused on the content in \ref{section2} as well as the development of the computer code for the Ostrogradsky algorithm. Mr. Pontius also focused on the content in \ref{section2}, as well as the example problems. Dr. Torre led the effort, providing guidance and feedback on all fronts. In the preparation of the manuscript, Mr. Watson and Mr. Pontius were responsible for the writing and editing of the manuscript, while Dr. Torre was responsible for drafting initial versions.

\bibliographystyle{apsrev4-2}
\bibliography{references}

\end{document}